\documentstyle[epsf,cite,12pt]{article}
\begin{document}
\newcommand{\ome}{\omega_{\rm rot}}
\newcommand{\sumA}{\sum_{i=1}^A}
\newcommand{\NZ}{$N$=$Z$~}
\newcommand{\mass}{$A$=30-50~}
\newcommand{\Su}{${}^{32}$S}
\newcommand{\Ar}{${}^{36}$Ar}
\newcommand{\Ca}{${}^{40}$Ca}
\newcommand{\Ti}{${}^{44}$Ti}
\newcommand{\Cr}{${}^{48}$Cr}
\newcommand{\boldr}{\mbox{\boldmath$r$}}
\newcommand{\boldj}{\mbox{\boldmath$j$}}
\newcommand{\boldrho}{\mbox{\boldmath$\rho$}}

\title{
Cranked Skyrme-Hartree-Fock calculation for superdeformed and
hyperdeformed rotational bands in \NZ nuclei from \Su~ to \Cr
}
\author{T. Inakura$^a$, S. Mizutori$^b$, M. Yamagami$^{a,c}$ 
and K. Matsuyanagi$^a$\\
{\small\it $^a$ Department of Physics, Graduate School of Science,}\\
{\small\it Kyoto University, Kitashirakawa, Kyoto 606-8502, Japan}\\
{\small\it $^b$ Department of Human Science, Kansai Women's College,}\\
{\small\it Kashiwara City, Osaka 582-0026, Japan}\\
{\small\it $^c$ Institut de Physique Nucl\'eaire, IN$_{2}$P$_{3}$-CNRS,}\\
{\small\it 91406 Orsay Cedex, France}
}
\date{}
\maketitle

\begin{abstract}

With the use of the symmetry-unrestricted cranked Skyrme-Hartree-Fock
method in the three-dimensional coordinate-mesh representation,
we have carried out a systematic theoretical search for the
superdeformed and hyperdeformed rotational bands 
in the mass \mass region. 
Along the \NZ line, we have found superdeformed solutions in 
~\Su, ~\Ar, ~\Ca, ~\Ti, 
and hyperdeformed solutions in ~\Ar, ~\Ca, ~\Ti, ~\Cr.   
The superdeformed band in ~\Ca~ is found to be extremely soft against 
both the axially symmteric ($Y_{30}$) and asymmetric ($Y_{31}$)
octupole deformations.
An interesting role of symmetry breaking in the mean field is pointed out.



\end{abstract}

\vspace{2cm}
\newpage
\section{Introduction}

Nowadays, about two hundreds superdeformed (SD) rotational bands are 
identified in various mass ($A$=60, 80, 130, 150, 190) regions
\cite{nol88,abe90,jan91,bak95,bak97,dob98}.
Every regions of superdeformation have their own
characteristics so that we can significantly enlarge and 
deepen our understanding of nuclear structure by systematically
investigating similarities and differences among the SD bands 
in different mass region.
For the mass \mass region,
although the doubly magic SD band in ~\Su, 
which has been expected quite a long time~
\cite{she72,lea75,rag78,ben81,gir83,mol00,rod00,tan01,afa00},
has not yet been observed and remains as a great challenge\cite{dob98},
quite recently, beautiful rotational
spectra associated with the SD bands have been observed up to high spin
in neighboring \NZ nuclei; ~\Ar, ~\Ca, and ~\Ti. 
In ~\Ar~ the SD band has
been identified up to its termination at $I^\pi=16^+$
\cite{sve00,sve01a,sve01b}. 
The SD band in the spherical magic nucleus \Ca~ is built on the well 
known 8p-8h excited $0^+$ states at 5.213 MeV 
and the rotational spectra have been observed 
up to $I^\pi=16^+$\cite{ide01}.
In ~\Ti~ a rotational spectrum associated with the excited $0^+$ state
at 1.905 MeV has been observed up to $I^\pi=12^+$\cite{lea00}. 
This rotational band may also be regarded as belonging to a family
of the SD band configurations.
The fact that rotational bands built on excited
$0^+$ states are systematically observed is a quite important, 
unique feature of the SD bands in the ~\Ca~ region, 
as the low angular momentum portions of the SD bands in heavier mass 
regions are unknown in almost all cases.

In nuclei along the \NZ line,  
effects of deformed shell structures of protons and neutrons 
act coherently and rich possibilities arise for coexistence and 
competition of different shapes. Thus, 
we shall be able to learn details of deformed shell structure 
and microscopic mechanism of shape coexistence by a systematic
study of high-spin yrast structure in the sequence of \NZ nuclei. 
Especially, yrast spectroscopy of nuclei in the \mass region,
being relatively light compared to other regions of SD nuclei, 
is expected to provide detailed information about the roles of
individual deformed single-particle orbits
responsible for the emergence of the SD bands.

In this paper, as a continuation of the previous work 
on ~\Su~\cite{yam00}, we carry out a systematic theoretical search
for SD and more elongated hyperdeformed (HD) rotational bands
in \NZ nuclei from ~\Su~ to ~\Cr~
by means of the symmetry-unrestricted, 
cranked Skyrme-Hartree-Fock (SHF) method.
In Ref. \cite{yam00}, a new computer code was constructed 
for the cranked SHF calculation based on the 
three-dimensional (3D) Cartesian-mesh representation, 
which provides a powerful tool for exploring exotic shapes 
(breaking both axial and reflection symmetries in the intrinsic states) 
at high spin. The algorithm of this code for numerical calculation 
is basically the same as in 
Refs.\cite{dav80,flo82,flo84,bon85,bon87,taj96,taj98,taj01,tak96,tak98},
except that various restrictions on spatial symmetries are completely
removed. Namely, we do not impose parity and signature symmetries 
on intrinsic wave functions. Hence we call this version of the
cranked SHF method ``symmetry-unrestricted" one.
For the development of selfconsistent mean-field models
for nuclear structure, 
we quote Refs. \cite{abe90}, \cite{rei99} and \cite{ben02},
in which various kinds of mean-field theory, including Hartree-Fock (HF)
calculations with finite-range Gogny interactions \cite{dec80}
and relativistic mean-field approaches \cite{rin96}, 
are thoroughly reviewd.
We also mention that spontaneous symmetry breaking in rotating nuclei
is reviewd in \cite{fra01}.

In fact, SD and HD solutions of the SHF equations we report in this paper
preserve the reflection symmetries with respect to the $(x,y)$, $(y,z)$
and $(z,x)$ planes, so that the symmetry-unrestricted calculation gives
identical results with those evaluated by imposing such symmtries.
The symmetry-unrestricted calculation, however, enables us 
to examine stabilities of the SD and HD states against such
reflection-symmetry breaking degrees of freedom like
octupole deformations. In addition, we shall show that the symmetry 
breaking play a quite interesting role in the crossing region between
different configurations  
away from the local minima in the deformation parameter space.

This paper is arranged as follows:
In Section~2,
a brief account of the cranked SHF method is given. 
In Section~3,
results of calculation for deformation energy curves 
and the SD and HD rotational bands 
in nuclei from ~\Su~ to ~\Cr~ are systematically presented. 
Here, special attention will be paied to the properties of the SD bands 
at their high spin limits and the crossover to the HD bands 
with increasing angular momentum. 
In Section~4,
an interesting role of symmetry breaking in the mean field 
will be pointed out in connection with 
configuration rearrangement mechanism.
We shall further make a detailed analysis of the SD band of ~\Ca~
and show that it is extremely soft against 
both the axially symmteric ($Y_{30}$) 
and asymmetric ($Y_{31}$) octupole deformations. 
Main results of this paper are summarized in Section~5.

A preliminary version of this work was reported in \cite{yam00b,ina01}.

\section{Cranked SHF calculation}

The cranked HF equation for a system uniformly rotating about the
$x$-axis is given by 

\begin{equation}
\delta<H - \ome J_x>=0,
\end{equation}

\noindent
where $\ome$ and $J_x$ mean the rotational frequency and
the $x$-component of angular momentum,  
and the bracket denotes the expectation value with respect to a
Slater determinantal state.
We solve the cranked HF equation for a Hamiltonian of the Skyrme type 
by means of the imaginary-time evolution technique\cite{dav80}
in the 3D Cartesian-mesh representation. 
We adopt the standard algorithm
\protect\cite{dav80,bon85,taj96,bon87} in the numerical calculation,
but completely remove various restrictions on spatial symmetries.

When we allow for the simultaneous breaking of both reflection and
axial symmetries, it is crucial to accurately
fulfill the center-of-mass condition

\begin{equation}
<\sumA x_i>=<\sumA y_i>=<\sumA z_i>=0,
\end{equation}

\noindent
and the principal-axis condition 

\begin{equation}
<\sumA x_iy_i>=<\sumA y_iz_i>=<\sumA z_ix_i>=0.
\end{equation}

\noindent
For this purpose we use
the constrained HF procedure with quadratic constraints
\cite{flo73}. Thus, we replace the ``Routhian''
$R=H - \ome J_x$ in Eq. (1) with

\begin{equation}
R' = R - \sum_{k=1}^3 \mu_k <\sumA (x_k)_i>^2 
       - \sum_{k<k'}^3 \mu_{k,k'} <\sumA (x_kx_{k'})_i>^2.
\end{equation}

\noindent
In numerical calculations, we confirmed that 
the constraints (2) and (3) are
fulfilled to the order $O(10^{-15})$ 
with values of the parameters $\mu_k \sim O(10^2)$ and 
$\mu_{k,k'} \sim O(1)$. 
We solved these equations inside the
sphere with radius $R$=10~fm and mesh size $h$=1~fm, starting 
with various initial configurations.
We note that the accuracy for evaluating deformation energies 
with this mesh size was carefully checked 
by Tajima {\it et al}.~\cite{taj96,taj01} 
(see also Ref. \cite{bay86}) and was found to be quite satisfactory.
The 9-point formula was used as the difference formula
for the Laplacian operator.
As usual, the angular momentum is evaluated as
\\ $I\hbar=<J_x>$.
For the Skyrme interaction, we adopt the widely used three versions;
SIII~\cite{bei75}, SkM$^*$\cite{bar82} and SLy4~\cite{cha98}.

In addition to the symmetry-unrestricted cranked SHF calculation explained
above, we also carry out, for comparison sake, symmetry-restricted
calculations imposing reflection symmetries about 
the $(x,y)$-, $(y,z)$- and $(z,x)$-planes. 
The computational algorithm for this restricted version of 
the cranked SHF calculation is basically the same as in \cite{bon87}, but 
we have constructed a new computer code for this purpose. 
Below we call these symmetry-unrestricted and -restricted cranked
SHF versions ``unrestricted" and ``restricted" ones, respectively. 
Comparison between results obtained by unrestricted and 
restricted calculations carried out independently serves as  
a check of numerical results to be presented below.
Physical significance of this comparison is, however, that we can,
in this way, clearly identify effects of symmetry-breaking 
in the mean field.
We shall indeed find an interesting symmetry breaking effect 
in the next section.
 
Solutions of the cranked SHF equation give minima in the deformation energy
surface. In order to explore the deformation energy surface around 
these minima and draw 
deformation energy curves as functions of deformation parameters, 
we carry out the constrained HF procedure 
with quadratic constraints\cite{flo73}.  
Namely, in addition to the constraints to fulfil the center-of mass
and principal-axis conditions mentioned above, we also introduce constraints
involving relevant mass-multipole moment operators and solve resulting
constrained HF equations.

As measures of the deformation, we calculate the mass-multipole moments,  

\begin{eqnarray}
\alpha _{lm}={{4\pi } \over {3A\bar{R}^l}}\int r^l X_{lm}\left( 
{\Omega} \right)\rho \left( \boldr \right)d \boldr
,\quad (m=-l,\cdots ,l)
\end{eqnarray}

\noindent
where 
$\rho(\boldr)$ is the density,
$\bar{R}=\sqrt{5<\sumA \boldr_i^2>/3A}$, 
and 
$X_{lm}$ are real bases of the spherical harmonics,

\begin{eqnarray}
	X_{l0} & = & Y_{l0},
		 \\
	X_{l|m|} & = & \frac{1}{\sqrt{2}}( Y_{l-|m|}+Y_{l-|m|}^{*} ),
		   \\
	X_{l-|m|} & = & \frac{-i}{\sqrt{2}} ( Y_{l|m|}-Y_{l|m|}^{*} ).
\end{eqnarray}
\noindent
Here the quantization axis is chosen as the largest (smallest) 
principal axis for prolate (oblate) solutions. 
We then define the quadrupole deformation parameter $\beta_2$,
the triaxial deformation parameter $\gamma$, and the
octupole deformation parameters $\beta_3$ and $\beta_{3m}$ by

\begin{eqnarray}
\alpha _{20}=\beta _2\cos \gamma ,
\quad \alpha _{22}=\beta _2\sin \gamma,
\end{eqnarray}
\begin{eqnarray}
\beta _3=\left( {\sum\limits_{m=-3}^3 {\alpha _{3m}^2}} \right)^{1/2},
\quad \beta _{3m}=\left( {\alpha _{3m}^2+\alpha _{3-m}^2} \right)^{1/2}
\quad \left( {m=0,1,2,3} \right). 
\end{eqnarray}

For convenience, we also use the familiar notation $-\beta_2$ for
oblate shapes with $(\beta_2,\gamma=60^\circ)$.

\section{Results of calculation}
\subsection{\it Deformation energy curves}

Figures 1-3 show deformation energy curves   
evaluated at $I=0$ by means of the constrained HF procedure
with the quadratic constraint on the mass-quadrupole moment.
The SIII, SkM$^*$, and SLy4 versions of the Skyrme interaction
are used in Figs.~1, 2, and 3, respectively.
Solid lines with and without filled circles in these figures 
represent results 
of unrestricted and restricted calculations, respectively.
Let us focus our attention to the region of large quadrupole
deformation $\beta_2$.
In both cases, 
we otain local minima corresponding to the SD states
for ~\Su, ~\Ar, ~\Ca~ and ~\Ti~ 
in the region $0.4 \le \beta_2 \le 0.8$.
(The local minimum in ~\Ti~ is triaxial, as shown in Fig.~10 below, 
i.e., it is situated away from the $\gamma=0$ section of the deformation 
energy surface, so that it is not clearly seen in Figs.~1-3.)
The local minima in ~\Su~ and ~\Ar~ involve four particles
(two protons and two neutrons)
in the $fp$ shell, while those in
~\Ca~ and ~\Ti~ involve eight particles (four protons and four neutrons).
These local minima respectively correspond to the 
4p-12h, 4p-8h, 8p-8h and 8p-4h configurations with respect to 
the doubly closed shell of ~\Ca, 
and their properties have been discussed 
from various point of view; see Refs.~
\cite{she72,lea75,rag78,ben81,gir83,mol00,rod00,tan01,afa00} for ~\Su,
Refs.~\cite{sve00,sve01a,sve01b,lon01} for ~\Ar,
Refs.~\cite{ide01,ger67,met71,ger77,zhe88,zhe90,kan01,cau02} for ~\Ca,
and Refs.~\cite{sim73,mic98,lea00} for ~\Ti.

In addition to these SD minima, 
we also obtain local minima 
in the region $\beta_2 \geq 0.8$ for ~\Ca, ~\Ti~ and ~\Cr. 
These minima involve additional four particles 
(two protons and two neutrons)
in the single-particle levels that reduces to the g$_{9/2}$ levels
in the spherical limit.
Somewhat loosely we call these local minima ``hyperdeformed."
The HD solution in ~\Ca~ corresponds to the 12p-12h configuration.
For \Ti, we obtain two HD solutions which correspond to 
the 12p-8h and 16p-12h configurations. 
The HD solution in ~\Cr~ corresponds to the 16p-8h configuration.
These HD solutions well agree with those previously obtained in
the SHF calculation by Zheng, Zamick and Berdichevsky~\cite{zhe90}. 
We also mention that the 12p-12h configuration in ~\Ca~
and the 16p-12h configuration in ~\Ti~ agree with those
obtained by the macroscopic-microscopic model calculation
by Leander and Larsson~\cite{lea75}.

In Figs.~1-3 and in the following, 
the above SD and HD configurations are denoted
by $f^n g^m$ (or $(fp)^n g^m$), 
where $n$ and $m$ indicate the numbers of nucleons occupying 
the $f_{7/2}$ shell (or the $fp$ shell) and the $g_{9/2}$ shell, 
respectively.

As seen in Figs.~1-3, these SD and HD minima are obtained for all 
calculations with the use of the SIII, SkM$^*$, and SLy4 interactions.
These local minima preserve the reflection symmetries so that the
results of restricted and unrestricted calculations are the same.
On the other hand, we also find a case where the two calculations 
give different results:
We obtain a HD minimum  
with $\beta_2 \simeq 0.8$ for ~\Ar~ in the restricted calculations.
This minimum involves eight particles (four protons and four neutrons)
in the $fp$ shell	
and correspond to the 8p-12h configuration, but it disappears
in the unrestricted calculations and its remnant remains as a shoulder
of the deformation energy curve. 

Although the restricted and unrestricted calculations give
identical results for the SD and HD local minima except for
the HD solution for ~\Ar, they show different behaviors 
in regions away from the local minima: 
In Figs. 1-3, we see that the
deformation energy curves obtained by the
unrestricted calculations always join different local minima smoothly.
On the other hand, in the restricted calculations, 
segments of the deformation energy curves associated
with different local minima sharply cross each other in some situations,
while they are smoothly joined in other situations.
Closely examining the configurations involved,
we notice that the sharp crossings occur between configurations
having different numbers of particles excited into the $fp$ shell.
This point will be further elaborated in the subsequent section.

In Figs.~1-3 there are a number of local minima in the region of
smaller values of $\beta_2$. We shall not discuss on these local minima
in this paper, since the pairing correlations not taken into account 
here are expected to be important for these.  

The single-particle level schemes at the SD and HD local
minima mentioned above are displayed in Figs.~4-6 
at positions of their quadrupole deformation parameters $\beta_2$. 
One may notice that 
these equilibrium deformations correspond to 
values of $\beta_2$ slightly smaller 
than those where
the amount of energy spacings between the highest occupied and
lowest unoccupied single-particles levels
become local maxima.
This is a characteristic common to three versions of the
Skyrme interactions used,
and in accord with the expectation that 
equilibrium deformations are determined by the
sum of microscopic (shell-structure) and macroscopic (liquid-drop)
energies; the latter shifts the equilibrium values of
$\beta_2$ to slightly smaller ones.

\subsection{\it SD and HD rotational bands}

Let us focus our attention to the SD and HD local minima 
shown in Figs.~1-3, 
and investigate properties of the rotational bands 
built on them. 
Figures 7-9 show excitation energies, as functions of angular momentum,
of the SD and HD rotational bands 
calculated with the use of the SIII, SkM$^*$, and SLy4 interactions,
respectively.
These rotational bands are obtained by cranking each SHF solution 
(the SD and HD local minima in Figs.~1-3) 
and following the same configuration with increasing value of $\ome$
until the point where we cannot clearly identify the continuation
of the same configuration any more. 
We note that, in ~\Ti, two HD bands associated with 
the $f^8g^4$ and $(fp)^{12}g^4$ configurations 
cross at $I = 30-34$, and the latter becomes 
the yrast for higher spin.
(This band continues beyond $I=40$ where the figure is cut.)
In addition to the SD and HD bands built on the 
$I^\pi=0^+$ band-head states, we have found a HD band
in ~\Ar, which does not exist at $I=0$ and emerges at $I \simeq 16$
due to the rotation alignment of the $g_{9/2}$ orbit. 
This HD band is denoted by $f^6g^2$ and is included in Figs.~7-9. 
A similar configuration, $f^4g^2$, was found for ~\Su~ 
in our previous calculation~\cite{yam00} and called ``HD-like."
This and analogous configurations in nuclei other than ~\Ar~
are not illustrated in Figs.~7-9 in order not to make
the figure too complicated
(drawing complete yrast spectra of individual nuclei is
not the major purpose of these figures). 

As is well known, 
according to the deformed harmonic-oscillator potential model,
$N=Z=$ 18 and 24 are magic numbers associated with
the HD shell structure with axis ratio $3:1$.
These HD states respectively correspond to 
the $f^4g^4$ and $f^{12}g^4h^4$ configurations in our notation, 
where $h$ denotes the level associated with the $h_{11/2}$ shell.
Microscopic structures of the HD solutions under discussion
are apparently different from these, however.
We also mention that	
the possible existence of HD rotational bands at high spin 
in ~\Ar~ and ~\Cr~ have been discussed 
in Refs.~\cite{rae92, zha94} from the viewpoint of 
the cranked cluster model.
The relationship between our solutions and their
solutions associated with cluster structure is not clear.

Calculated quadrupole deformation parameters $(\beta_2,\gamma)$ 
of all bands mentioned above and their variations   
are displayed in Fig.~10.
The rotational frequency dependence of the 
single-particle energy levels (Routhian) is illustrated in Fig.~11,
taking the SD band in ~\Ca~ as a representative case.
The excitation energies of the SD and HD bands 
obtained by using different versions (SIII, SkM$^*$, SLy4)
of the Skyrme interaction are compared in Fig.~12 with
the experimental data \cite{sve00,ide01,lea00}.

Examining these figures, we see that, 
aside from quantitative details and 
some subtle points to be discussed below, 
the results obtained by using different versions of the Skyrme interaction
are similar. This implies that the basic properties of the SD and HD bands
under discussion are not sensitive to the details of the effective 
interaction.

As mentioned in the introduction, one of the unique features 
of the SD bands in the ~\Ca~ region is the possibility to observe
the SD rotational level structure from the $I^{\pi}=0^+$ 
band heads up to the maximum angular momenta allowed for 
the many-particle-many-hole configurations characterizing the
internal structures of these bands.
In fact, such a ``SD band termination" has been observed at $I=16$
in ~\Ar~ and well described by calculations 
in terms of the $j$-$j$ coupling shell model, 
the cranked Nilsson-Strutinsky
model\cite{sve00,sve01a,sve01b},
and the projected shell model\cite{lon01}.
On the other hand, for ~\Ca~ and ~\Ti, it is not clear whether or not 
the SD band continues beyond the highest spin states observed
up to now 
(the $16^+$ state in \Ca~\cite{ide01} and the $12^+$ state
in ~\Ti~\cite{lea00})
and, quite recently, their properties, from the $0^+$ band-heads to 
such high-spin regions, have been discussed
in terms of the spherical shell model in Ref.~\cite{cau02}
for ~\Ca~ and in Ref.~\cite{lea00} for ~\Ti.
In our calculation, 
except for the case of using the SLy4 interaction,
the band termination phenomenon in ~\Ar~ is reproduced;
the shape becomes triaxial and evolves toward the oblate shape,
although the oblate limit is not reached. 
In the cases of ~\Ca, the shape is slightly triaxial 
with $\gamma = 6^{\circ}$-$9^{\circ}
(8^{\circ}$-$9^{\circ})$
and the SD band terminates at $I \simeq 24$ 
in the calculation with the use of the SIII (SkM$^*$) interaction.
In the case of ~\Ti, the shape is more triaxial 
with $\gamma = 18^{\circ}$-$25^{\circ}$ and
$13^{\circ}$-$19^{\circ}$,
and the SD bands terminates at $I \simeq 12$ and 16
for the SIII and SkM$^*$ interactions, respectively.
Thus, the band termination properties appear quite 
sensitive to the details of the effective interaction.
Concerning the SD band termination in ~\Ca~ and ~\Ti,
the results obtained with the use of the SIII and SkM$^*$ interactions
would be more reliable than that with SLy4, 
in view of the above discussion for ~\Ar.
In any case, it would be very interesting to explore higher spin 
members of the SD rotational bands in ~\Ca~ and ~\Ti~ 
in order to understand the terminating properties 
of the SD bands at high spin limits.

As is clear from the comparison with experimental data in Fig.~12, 
the moments of inertia for the SD band 
are somewhat overestimated in the present calculation. 
To investigate a possible cause of this,
we plan to take into account the pairing correlations by means of 
the cranked Skyrme-Hartree-Fock-Bogoliubov code 
constructed in Ref.~\cite{yam01}.
One also notice that the excitation energy of 
the SD band-head state in ~\Ca~ is overestimated. 
It will decrease if the zero-point rotational energy correction,
$-\frac{1}{2{\cal J}}<J_x^2>$, 
is taken into account(see Ref. \cite{zhe90} for numerical examples).
Although the calculation of this correction is rather easy,
we need to evaluate, for consistency, also 
the zero-point vibrational energy corrections~\cite{rei99}, 
and this is not an easy task.
We therefore defer this task for a future publication.
Inclusion of these correlations is expected to improve
agreement with the experimental data.

\section{Discussions}

\subsection{\it A role of symmetry breaking}

Let us now discuss on the significance of 
the reflection symmetry breaking in the mean field.
As noticed in Figs.~1-3, the crossings between configurations involving
different numbers of particles in the $fp$ shell are sharp in the
restricted calculation, while we always obtain smooth configuration 
rearrangements in the unrestricted calculations.
The reason for this different behavior between the unrestricted and
restricted calculations is rather easy to understand:
When the parity symmetry is imposed, there is no way, within the mean-field
approximation, to mix configurations having different number of particles 
in the $fp$ shell.
In contrast, smooth crossover between these different  
configurations is possible via mixing between positive- and 
negative-parity single-particle levels, 
when such a symmetry restriction is removed. 
Let us examine this idea in more detail.  
In Figs.~13-15 octupole deformation parameters $\beta_3$ 
of the lowest energy states
for given values of $\beta_2$ are shown. 
They are obtained by the unrestricted SHF calculations
and plotted as functions of $\beta_2$ in the lower 
portion of each panel.
We see that $\beta_3$ are zero near the local minima in the
deformation energy surface, but rise in the crossing region
between configurations involving different number of particles
in the $fp$ shell. This means that the configuration
rearrangements in fact take place through paths in the deformation space 
that break the reflection symmetry. 
The importance of allowing the mean field for breaking 
symmetries in the process of configuration rearrangements 
was previously emphasized by Negele\cite{neg89} in their  
calculations for spontaneous fission of ~\Su~ by means of the
imaginary time tunneling method.

In connection with the above finding, it may be appropriate 
to point out another situation in which a symmetry breaking 
in the mean field plays an important role.   
For both the restricted and unrestricted 
calculations, we have obtained smooth crossover between the SD and HD
configurations in ~\Ca~ and ~\Ti~ (see Figs. 1-3). 
Since four particles are further excited into the $g$ shell in the HD
configurations, the smooth configuration rearrangement becomes possible
by means of the mixing between the down-sloping levels steming from the
$g_{9/2}$ shell (its asymptotic quantum number is $[440]\frac{1}{2}$)
and the up-sloping levels stemming from the $sd$ shell
($[202]\frac{5}{2}$ and $[200]\frac{1}{2}$ 
in the cases of ~\Ca~ and ~\Ti, respectively).
The mixing between these single-particle levels takes place through
the hexadecapole components of the mean field, and 
we need to break the axial symmetry to mix them in the case of ~\Ca.
The calculations called "restricted" in this paper allow the 
axial symmetry breaking, so that the smooth rearrangement 
from the SD to HD configurations is possible also in ~\Ca. 
A very careful computation is required, however, in order to
detect these mixing effects, since the interaction between the
down-sloping and up-sloping levels is extremely weak.

In Figs.~13-15, one may notice that 
$\beta_3$ take non-zero values also in some situations 
other than the crossing regions.
Such situations occur in some regions of the deformation energy surface
where it becomes very soft with respect to the reflection-asymmetric 
degrees of freedom.
In the next subsection, we investigate this point in detail 
taking the SD solution in ~\Ca~ as an especially interesting example.

\subsection{\it Octupole softness of the SD band in ${}^{40}$Ca}

Let us examine stabilities of the SD local minimum in ~\Ca~ 
against octupole deformations.
Figure 16 shows deformation energy curves as functions of 
the octupole deformation parameters  $\beta_{3m}(m=0,1,2,3)$  
for fixed quadrupole deformation parameters at and near the SD minimum
of ~\Ca, calculated by means of the constrained HF procedure 
with the use of the SIII, SkM$^*$, and SLy4 interactions.
We immediately notice that the SD state is extremely soft 
with respect to the $\beta_{30}$ and $\beta_{31}$ deformations,
irrespective of the Skyrme interactions used.
Although it is barely stable with respect to these directions
(see curves for $\beta_2=0.6$),
an instability toward the $\beta_{31}$ deformation occurs 
as soon as one goes away from the local minimum point
(see curves for $\beta_2=0.5$).
In fact, the deformation energy surface is found to be almost flat for 
a combination of the $\beta_{30}$ and $\beta_{31}$ deformations
already at the SD local minimum.
Thus we need to take into account the octupole shape fluctuations 
for a better description of the SD rotational band in ~\Ca.
It will be a very interesting subject to search for negative-parity
rotational bands associated with octupole shape fluctuation modes
built on the SD yrast band. We plan to make such a study in future. 
Quite recently, the octupole instability of the SD band in ~\Ca~ has 
been suggested also by Kanada-En'yo\cite{kan01}. 

\section{Conclusions}

With the use of the symmetry-unrestricted cranked SHF
method in the 3D coordinate-mesh representation,
we have carried out a systematic theoretical search for the
SD and HD rotational bands in the \NZ nuclei from ~\Su~ to ~\Cr.
We have found the SD solutions in ~\Su, ~\Ar, ~\Ca, ~\Ti,~ 
the HD solutions in ~\Ar, ~\Ca, ~\Ti, ~\Cr, and we have carried out
a systematic analysis of their properties at high spin.
 
It is explicitly shown that the crossover between configurations 
involving different number of particles in the $fp$ shell 
takes place via a reflection-symmery breaking path in the 
deformation space.

Particular attention has been paid to the recently discovered SD
band in ~\Ca, and 
we have found that the SD band in ~\Ca~is extremely soft against 
both the axially symmetric ($Y_{30}$) and 
asymmetric ($Y_{31}$) octupole deformations.
Thus, it will be very interesting to search for negative-parity
rotational bands associated with octupole shape vibrational
excitations built on the SD yrast band.

\section*{Acknowledgements}
We would like to thank E. Ideguchi, M. Matsuo, Y.R. Shimizu
and K. Hagino for useful discussions.
The numerical calculations were performed on the NEC SX-5 supercomputers
at RCNP, Osaka University, 
and at Yukawa Institute for Theoretical Physics, Kyoto University.
This work was supported by the Grant-in-Aid  for Scientific
Research (No. 13640281) from the Japan Society 
for the Promotion of Science.



\newpage

\begin{figure}
\epsfxsize=7 cm
\centerline{\epsffile{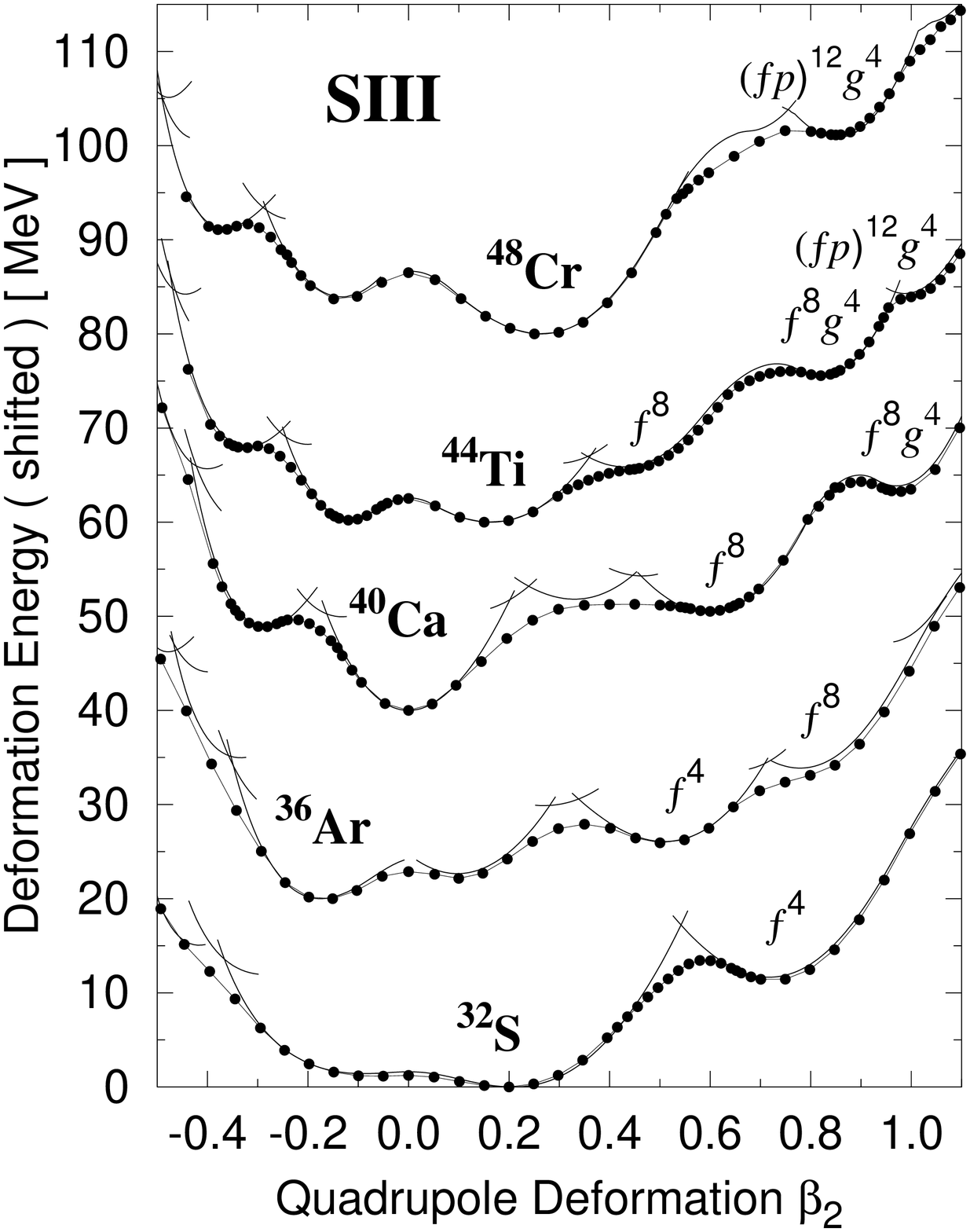}}
\caption{\small
Deformation energy curves as functions of the quadrupole deformation 
$\beta_2$ 
calculated at $I=0$ by means of the constrained SHF procedure with
the SIII interaction. 
The axial-asymmetry parameter $\gamma$ is constrained to be zero.
The curves for different nuclei are shifted 
by 20 MeV to accommodate them in a single plot.
Solid lines with and without filled circles represent 
the results obtained by the
unrestricted and restricted versions, respectively (see the text).
The notations $f^n g^m$ and $(fp)^n g^m$ indicate 
the configurations in which 
the $f_{7/2}$ shell ($fp$ shell) and the $g_{9/2}$ shell
are respectively occupied by $n$ and $m$ nucleons.
}
\end{figure}

\begin{figure}
\epsfxsize=7 cm
\centerline{\epsffile{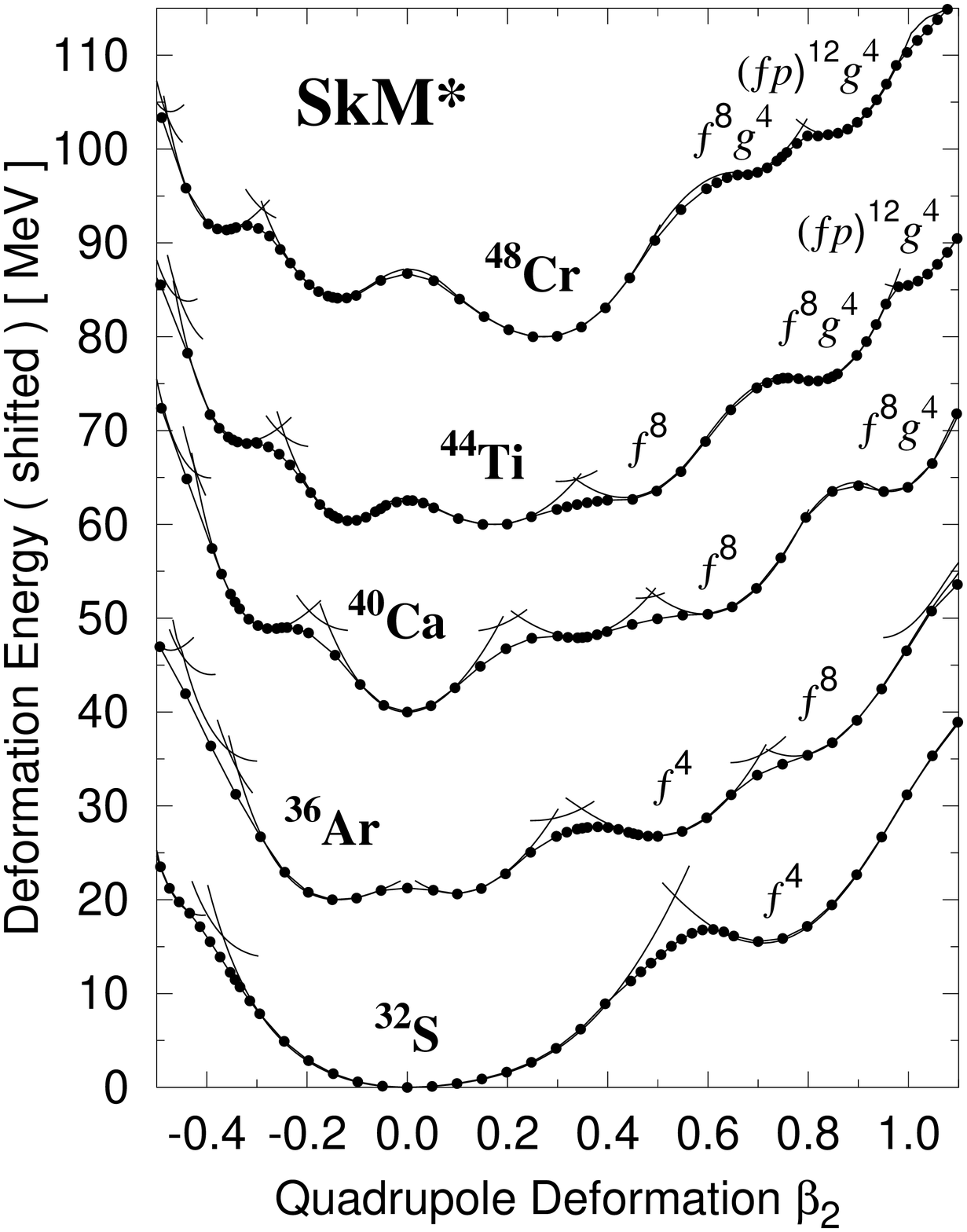}}
\caption{\small 
The same as Fig.1 but for the SkM$^*$ interaction.
}
\end{figure}

\begin{figure}
\epsfxsize=7 cm
\centerline{\epsffile{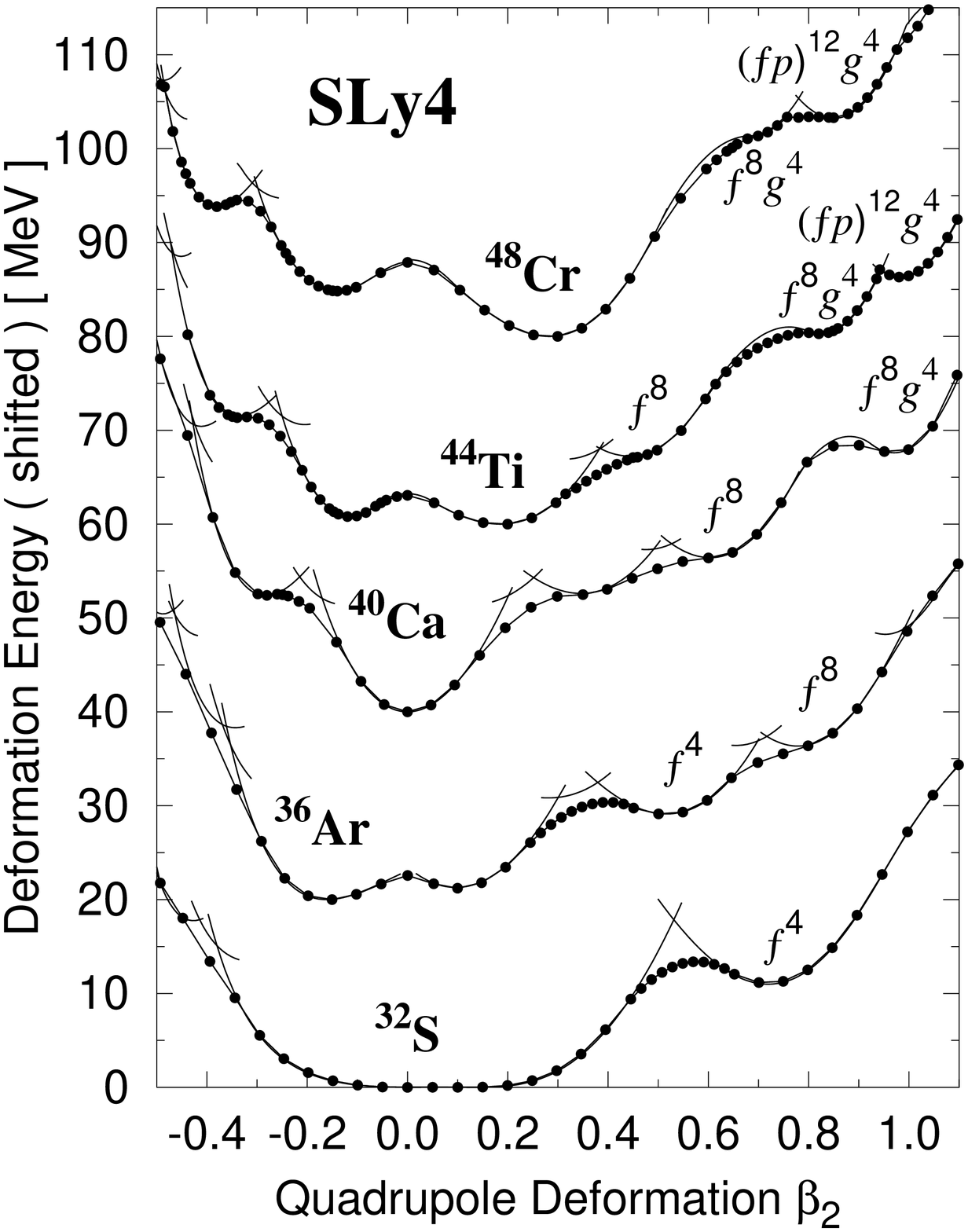}}
\caption{\small
The same as Fig.1 but for the SLy4 interaction.
}
\end{figure}

\begin{figure}
\epsfxsize=7 cm
\centerline{\epsffile{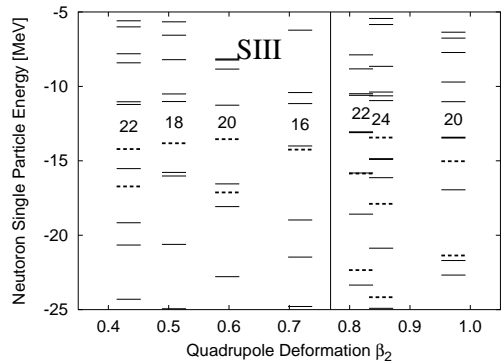}}
\caption{\small
Single-particle energy spectra for neutrons 
at various SD and HD local minima,
calculated at $I=0$ with the SIII interaction.
For every local minima, levels are drawn at positions
corresponding to their equilibrium values of $\beta_2$.
The levels are occupied up to the Fermi levels; 
on top of them, neutron numbers of the system are indicated.
The levels associated with the $f_{7/2}$ and $g_{9/2}$ shells
are drawn by broken and thick solid lines, respectively.
Single-particle spectra for protons are almost the same as 
those for neutrons.
}
\end{figure}

\begin{figure}
\epsfxsize=7 cm
\centerline{\epsffile{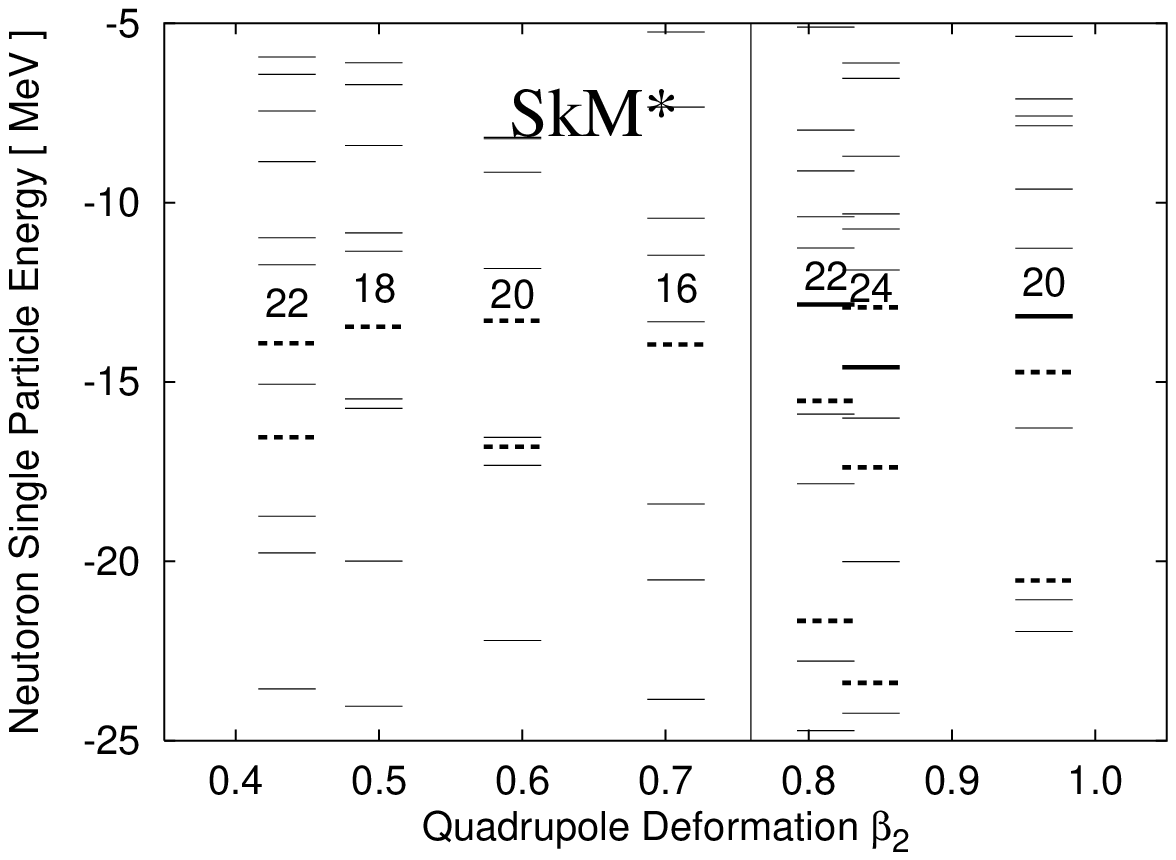}}
\caption{\small
The same as Fig.4 but for the SkM$^*$ interaction.
}
\end{figure}

\begin{figure}
\epsfxsize=7 cm
\centerline{\epsffile{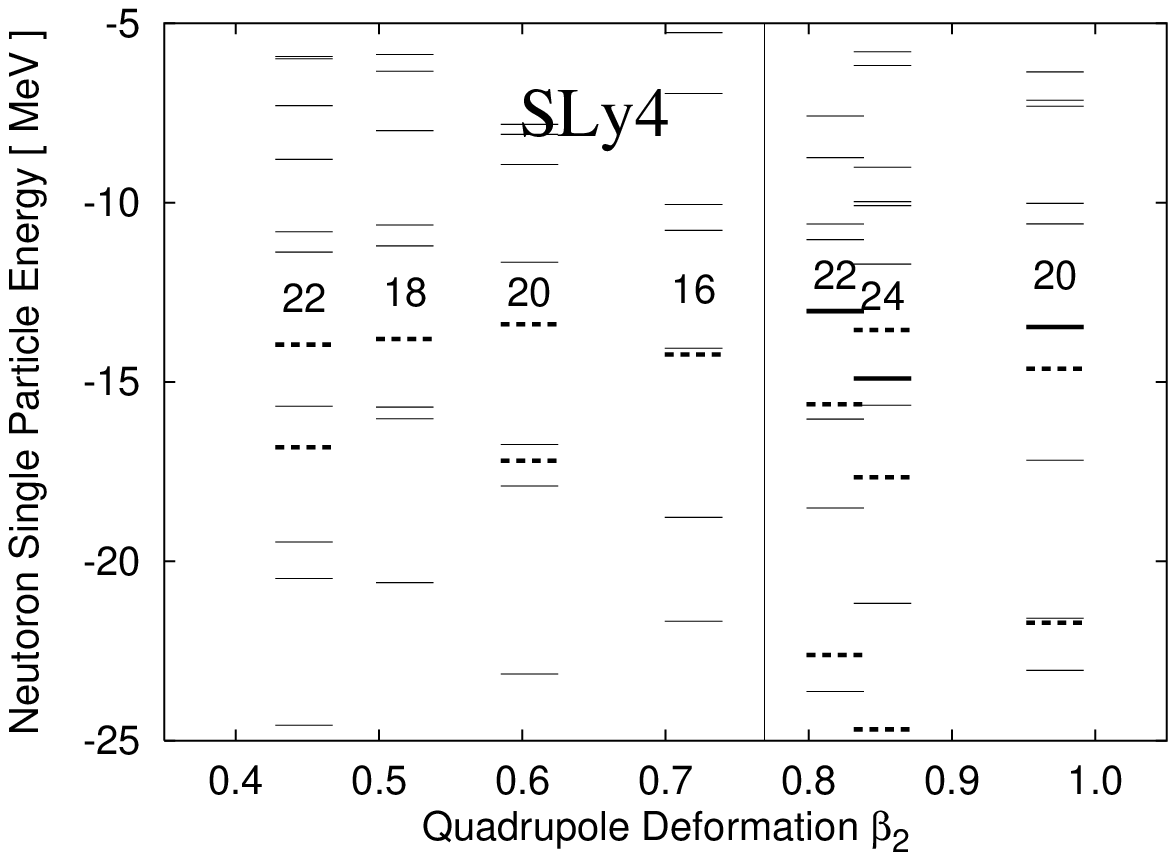}}
\caption{\small
The same as Fig.4 but for the SLy4 interaction.
}
\end{figure}

\begin{figure}
\epsfxsize=7 cm
\centerline{\epsffile{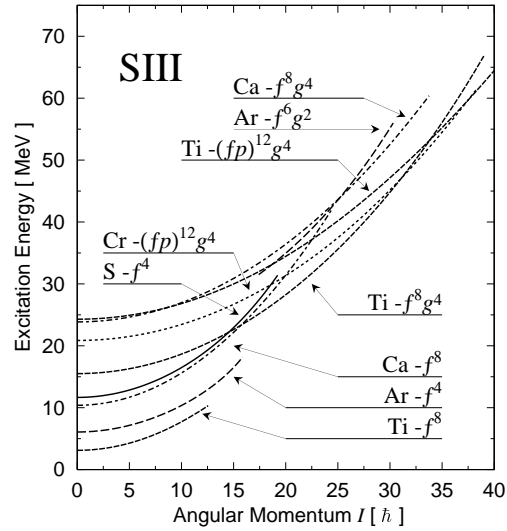}}
\caption{\small
Excitation energy vs. angular-momentum plot
for the SD and HD rotational bands obtained 
by the cranked SHF calculations with the use of the SIII interaction.
Their configurations are indicated by the same notations as in Fig.~1.
}
\end{figure}

\begin{figure}
\epsfxsize=7 cm
\centerline{\epsffile{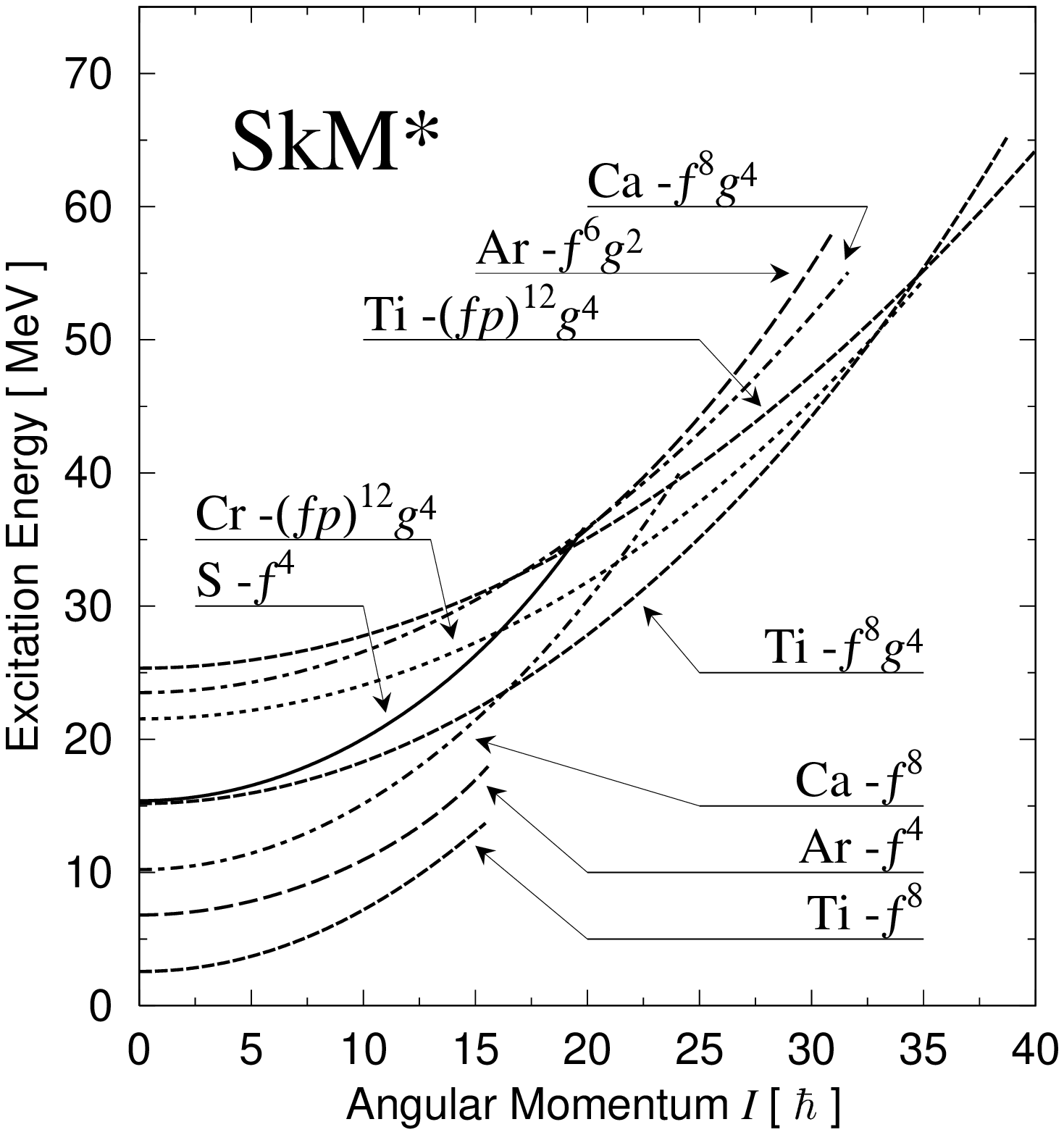}}
\caption{\small
The same as Fig.7 but for the SkM$^*$ interaction.
}
\end{figure}

\begin{figure}
\epsfxsize=7 cm
\centerline{\epsffile{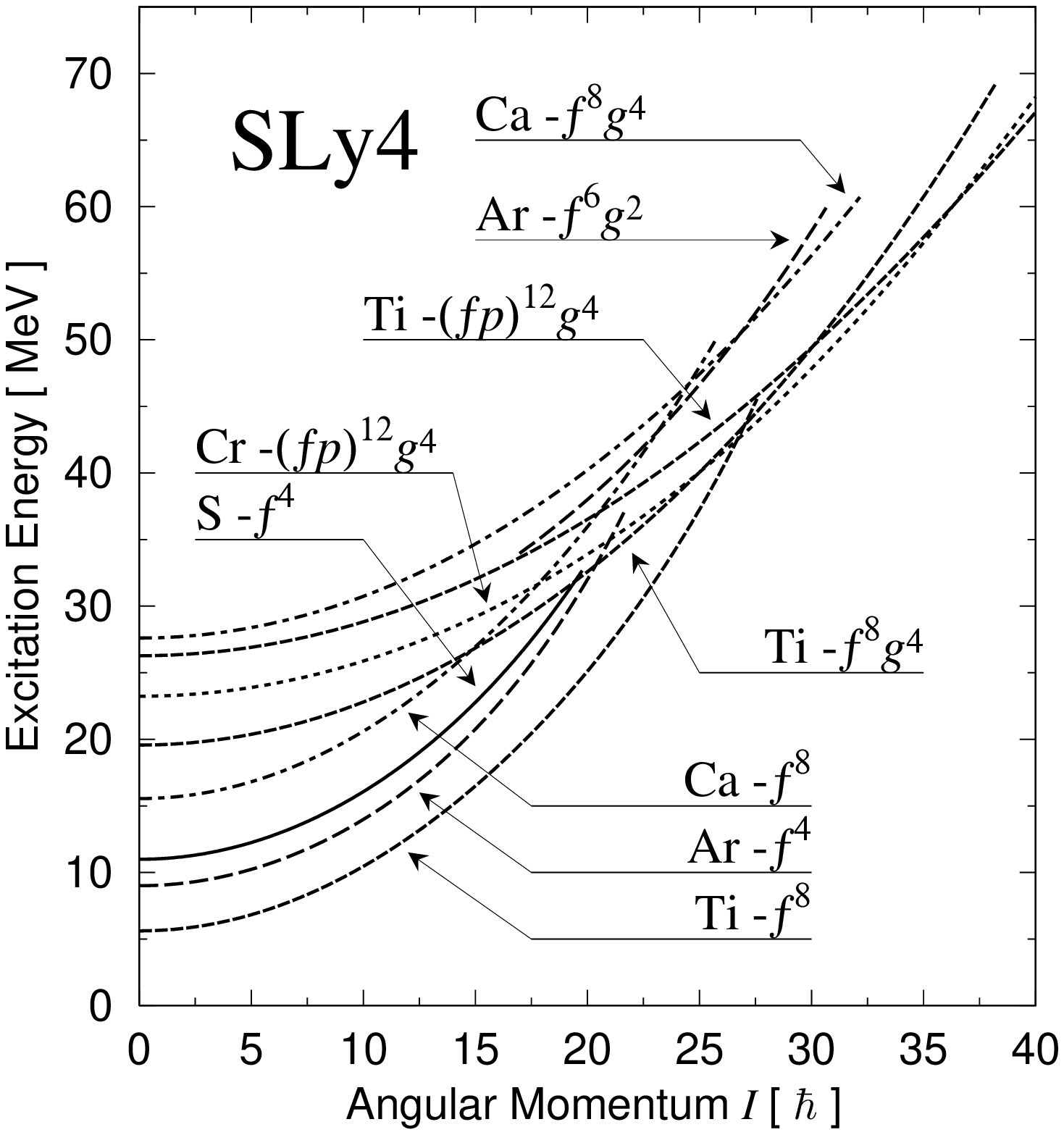}}
\caption{\small
The same as Fig.7 but for the SLy4 interaction.
}
\end{figure}

\begin{figure}
\epsfxsize=7 cm
\centerline{\epsffile{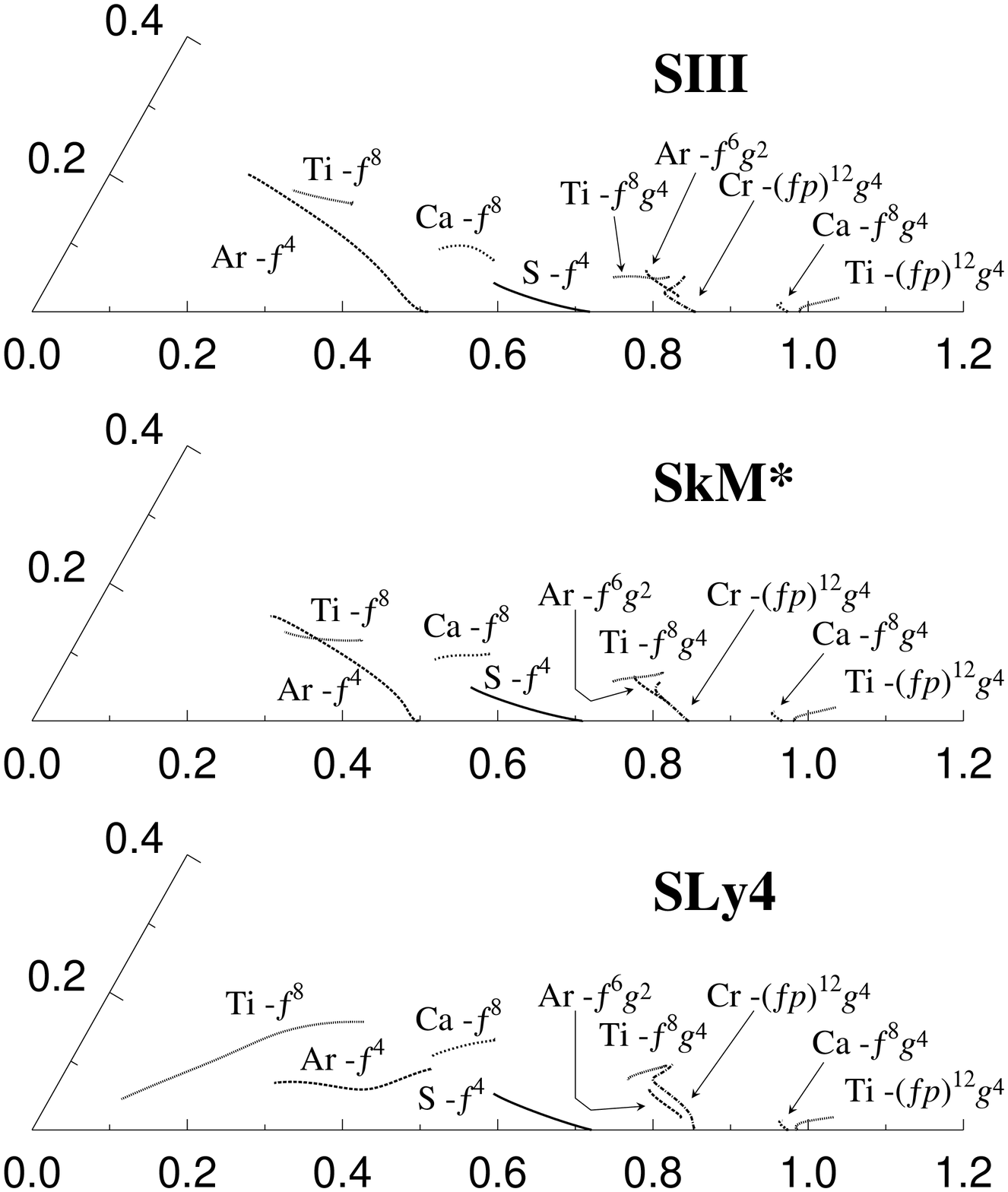}}
\caption{\small
Quadrupole deformation parameters $(\beta_2,\gamma)$ and
their variations  
for the SD and HD rotational bands in ~\Su,~\Ar,~\Ca,~\Ti~ and ~\Cr.
The top, middle and bottom panels show the shape evolution
in the $(\beta_2,\gamma)$ plane, evaluated with the
use of the SIII, SkM$^*$, and SLy4 interactions, respectively.
Notations and ranges of $I$ for individual bands are
the same as those shown in Figs.~7-9.
Values of $\beta_2$ decrease with increasing $I$, except for
the $(fp)^{12}g^4$ configuration in ~\Ti~(where $\beta_2$ increases with
increasing $I$).
}
\end{figure}

\begin{figure}
\epsfxsize=7 cm
\centerline{\epsffile{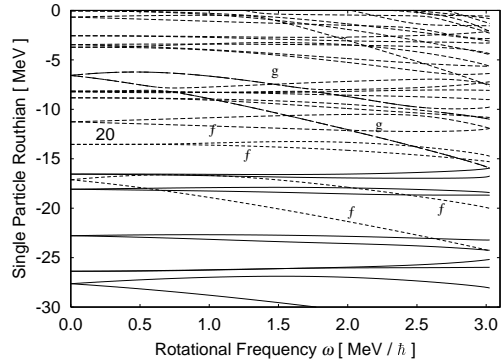}}
\caption{\small
Neutron single-particle energy diagram (Routhian) for the SD band in ~\Ca,
calculated with the use of the SIII interaction and plotted
as functions of rotational frequency $\ome$.
The levels associated with the $f_{7/2}$ and $g_{9/2}$ shells
are drawn by thick-broken and long-dashed lines, respectively.
}
\end{figure}

\begin{figure}
\epsfxsize=7 cm
\centerline{\epsffile{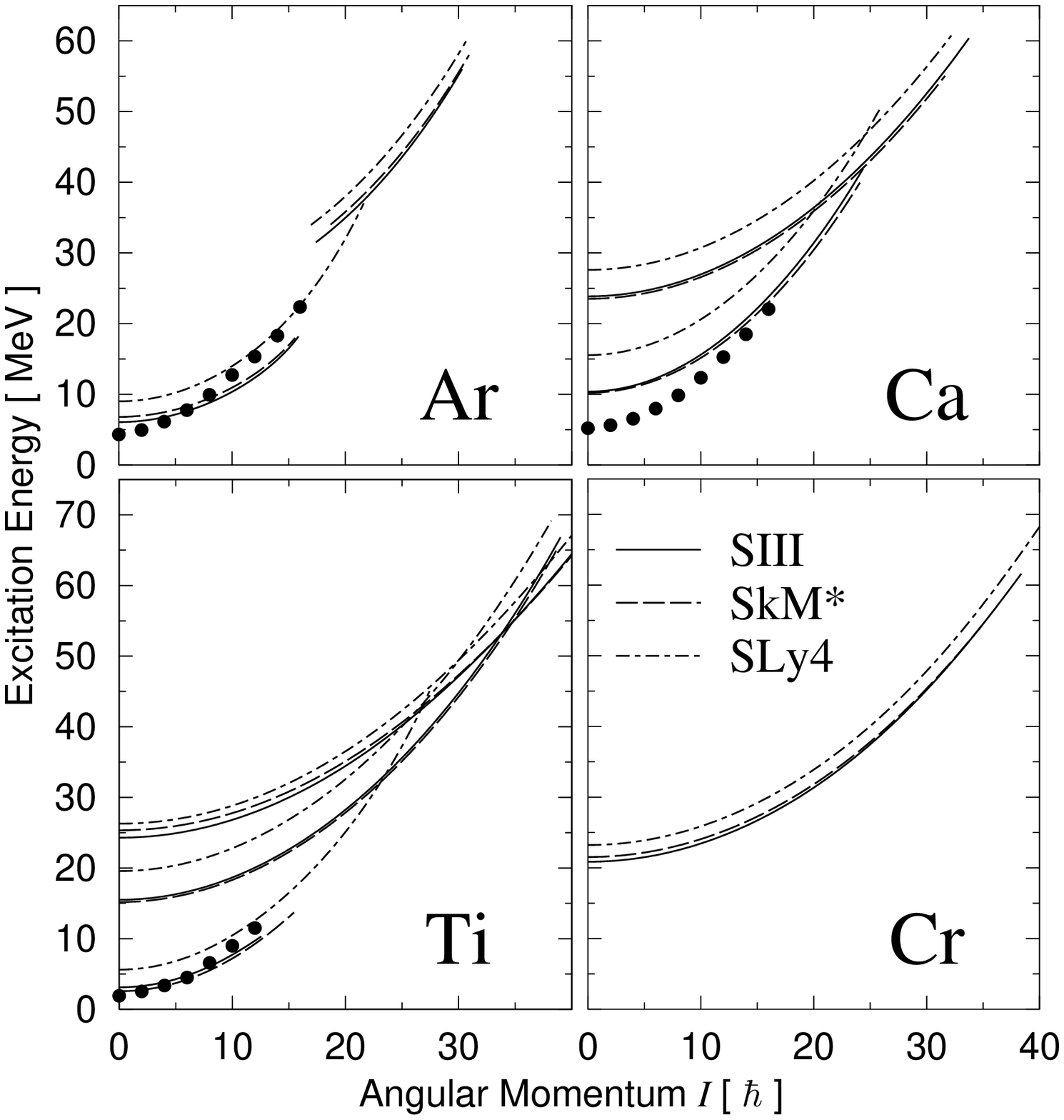}}
\caption{\small
Comparison between the excitation energies of the SD and HD bands 
in ~\Ar,~\Ca,~\Ti~ and ~\Cr, 
calculated by using different versions 
of the Skyrme interaction and 
the experimental data (\cite{sve00} for ~\Ar,~\cite{ide01} for ~\Ca,
and \cite{lea00} for ~\Ti).
The data are shown by filled circles and the results with
the SIII, SkM$^*$ and SLy4 interactions are drawn by solid, 
dashed and dashed-dotted lines, respectively.
}
\end{figure}

\begin{figure}
\epsfxsize=7 cm
\centerline{\epsffile{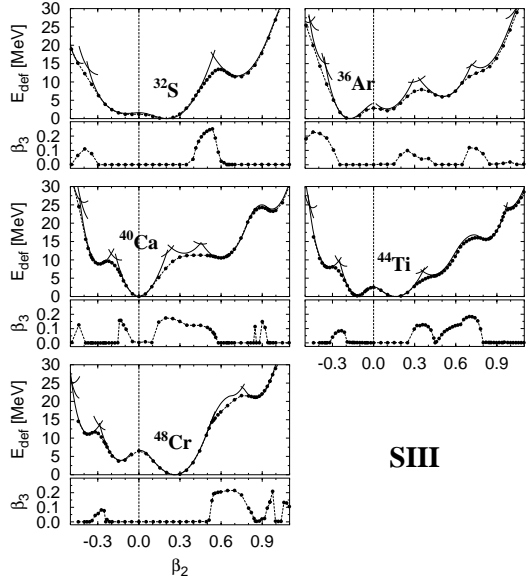}}
\caption{\small
Octupole deformation paramaters $\beta_3$ of the lowest energy states
for given values of $\beta_2$, obtained by the unrestricted SHF calculation
at $I=0$ with the use of the SIII interaction.
Their values are plotted 
as functions of $\beta_2$ in the lower portion of each panel.
To show that their values increase at crossing regions between 
configurations involving different number of particles 
in the $fp$ shell, the deformation energy curves are also displayed
in the upper portion of each panel. The latter are the same as those
presented in Fig.~1. 
}
\end{figure}

\begin{figure}
\epsfxsize=7 cm
\centerline{\epsffile{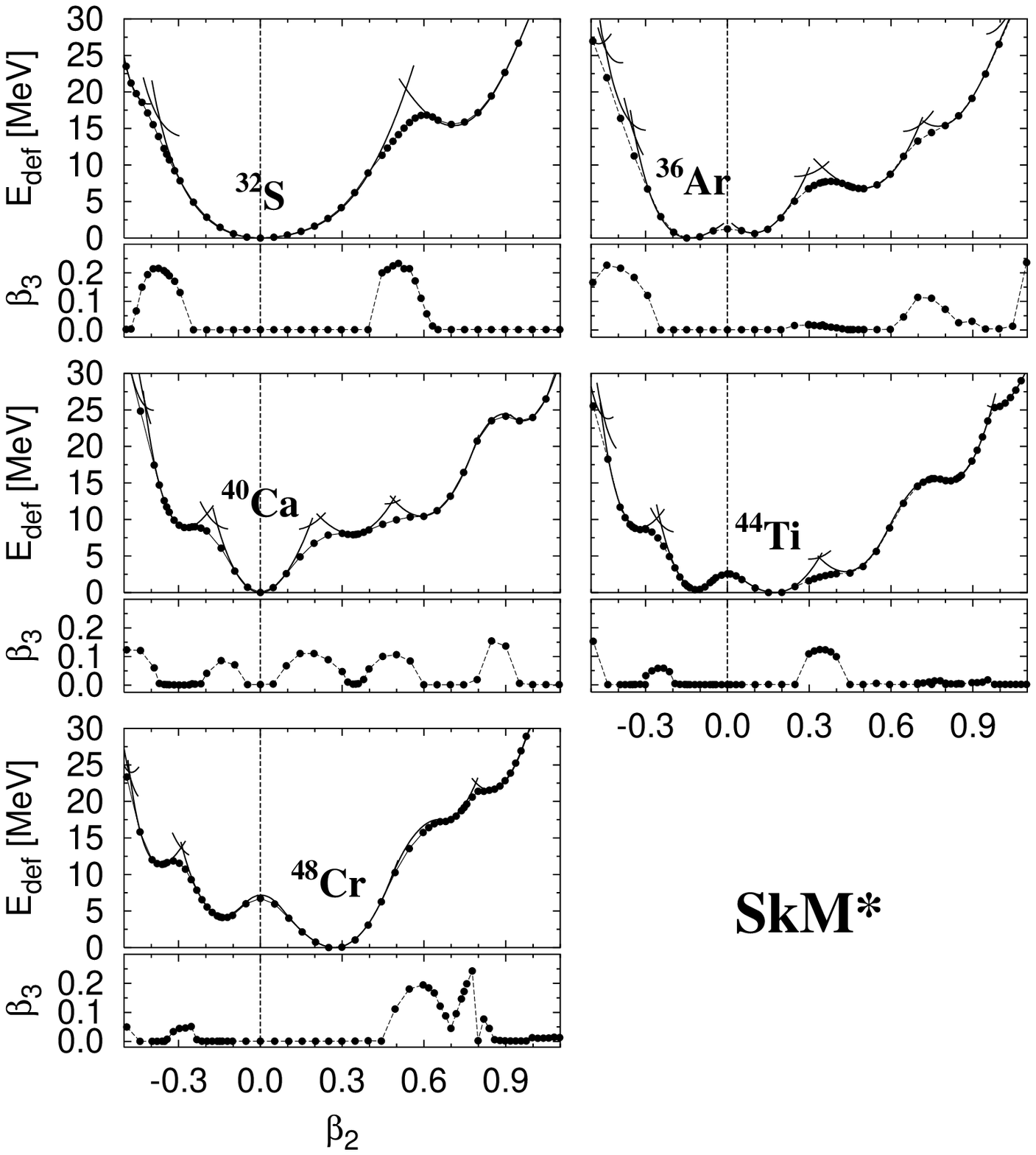}}
\caption{\small
The same as Fig.13 but for the SkM$^*$ interaction.
}
\end{figure}

\begin{figure}
\epsfxsize=7 cm
\centerline{\epsffile{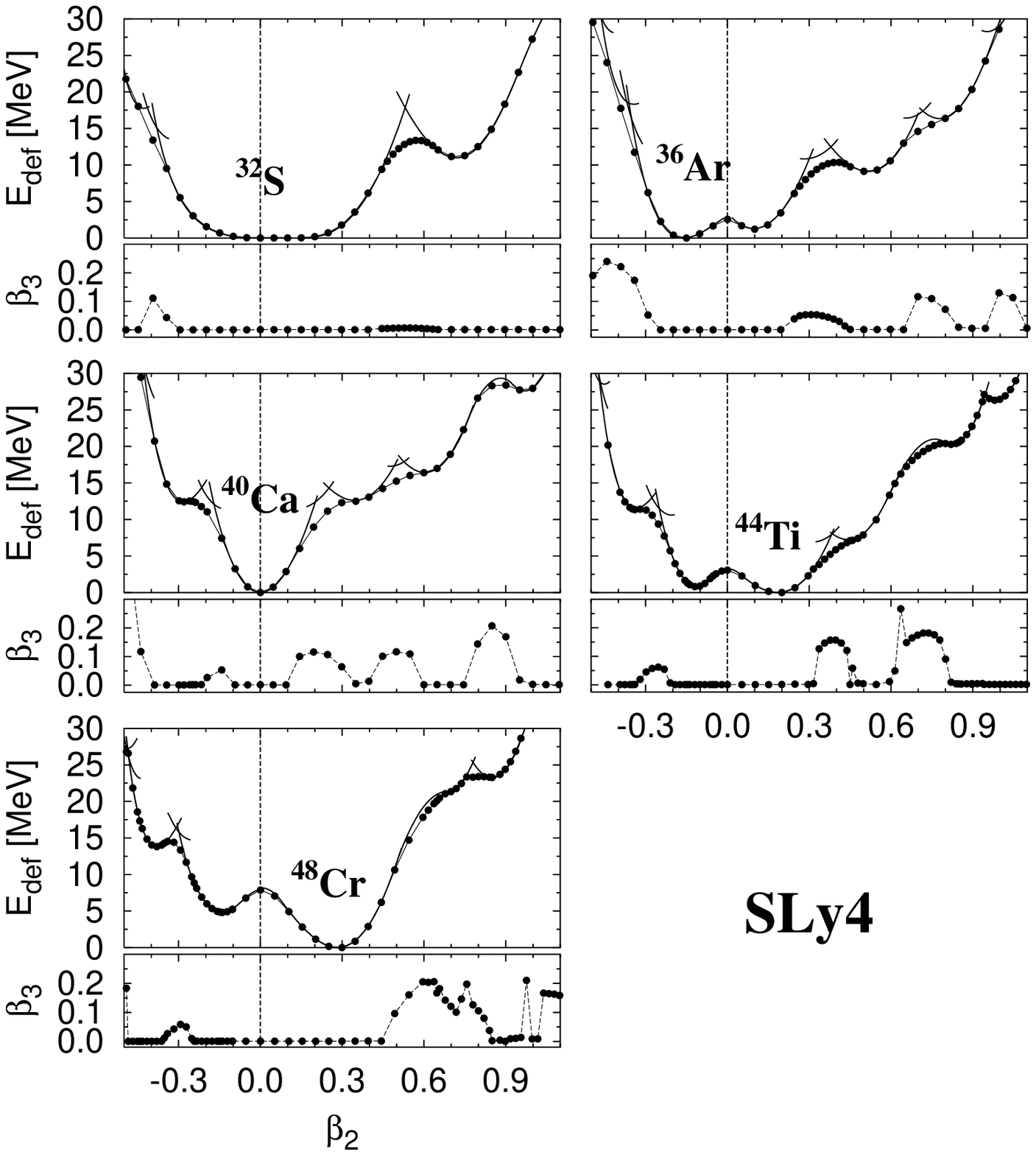}}
\caption{\small
The same as Fig.13 but for the SLy4 interaction.
}
\end{figure}

\begin{figure}
\epsfxsize=7 cm 
\centerline{\epsffile{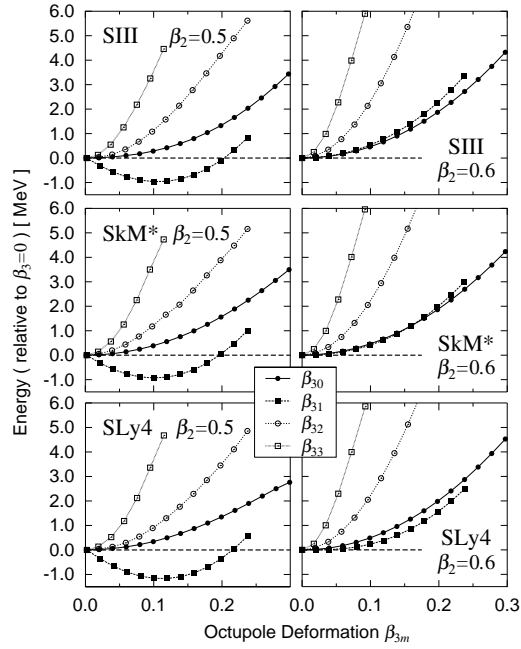}}
\caption{\small
Left-hand side: Deformation energy curves 
(measured from energies at $\beta_3=0$) as functions of 
the octupole deformation parameters  $\beta_{3m}(m=0,1,2,3)$,  
calculated for ~\Ca~ by means of the constrained HF procedure.
The quadrupole deformation parameters are fixed at $\beta_2=0.5$ 
and $\gamma=0$.
One of the $\beta_{3m}(m=0,1,2,3)$ is varied while the other
$\beta_{3m}$'s are fixed to zero.
Right-hand side: The same as the left-hand side, 
except that the quadrupole deformation parameters are fixed 
at $\beta_2=0.6$ and $\gamma=0$.
Results of calculation with the use of the SIII, SkM$^*$, and
SLy4 interactions are displayed in the upper, middle and lower panels,
respectively.
}
\end{figure}

\end{document}